# Analysis of Emission Dynamics of a Long Lifetime in Single InAs/GaAs Quantum Dots

Junhui Huang,[†,‡] Hao Chen,[†,‡] Zhiyao Zhuo,[†,‡] Jian Wang,[†,‡] Shulun Li,[†,‡] Kun Ding,[†] Haiqiao Ni,[†,‡] Zhichuan Niu,[†,‡,§] Desheng Jiang,[†] Xiuming Dou,*[,†,‡] and Baoquan Sun*[,†,‡,§]

[†] State Key Laboratory of Superlattices and Microstructures, Institute of Semiconductors, Chinese Academy of Sciences, Beijing 100083, China

[‡] College of Materials Science and Optoelectronic Technology, University of Chinese Academy of Sciences, Beijing 100049, China

[§] Beijing Academy of Quantum Information Sciences, Beijing 100193, China

**ABSTRACT**: A very long lifetime emission with non-single exponential decay characteristic has been reported for single InAs/GaAs quantum dot (QD) samples, in which there exists a long-lived metastable state in the wetting layer (WL) [ACS Photonics 2020,7,3228-3235]. In this article we have proposed a new three-level model to simulate the emission decay curve. In this model, assuming that the excitons in metastable state will diffuse and be trapped by QDs, and then emit fluorescence in QDs, a stretched-like exponential decay formula is derived as $I(t) = A\, t^{\beta-1} e^{-(rt)^\beta}$, which can well describe the long lifetime decay curve with an analytical expression of average lifetime $<\tau> = \frac{1}{r}\Gamma\left(\frac{1}{\beta}+1\right)$, where $\Gamma$ is the Gamma function. Furthermore, based on the proposed three-level model, an expression of the second-order auto-correlation function $g^2(t)$ which can well fit the measured $g^2(t)$ curve is also obtained.

By changing the emitter's electromagnetic local density of states (LDOS), the spontaneous radiative rate of the emitter will be changed as first pointed out by Purcell[1]. In the vicinity of the metal nanostructures, a modification of LDOS can be realized by confining the incident light within regions far below the diffraction limit to generate modes of localized surface plasmons (LSPs)[2-4], in which the spontaneous recombination rate is enhanced owing to the increase of LDOS. On the other hand, the recombination rate of emitter was reported to be reduced for an emitter in front of a planar interface[5], or near metal nanostructures[6]. This phenomenon is related to the destructive interference of the dipole fields between the emitters and reflected field, which results in a decrease of LDOS at the site of emitters.

In our previous report, a very long-lifetime of non-single exponential decay characteristic in InAs/GaAs single quantum dots (QDs) was observed due to the destructive interference of the dipole fields between the emitters (excitons in the wetting layer (WL)) and metal islands, which results in a decrease of the LDOS in two-dimensional excitons and the formation of a long-lifetime metastable state in the WL[6]. In fact, similar long lifetime with non-single exponential decay had been reported in many different systems, and for interpreting the long lifetime decay characteristics, a model with three-level configuration in which a third lower level of long-lived energy state was proposed. For example, a three-level charge-separated (CS) dark state model associated with quantum tunneling was applied successfully[7]. In addition, there were the log-normal distribution model assuming statistical law of rate distribution,[8] and the stretched exponential model caused by hopping-transport[9] or trapping[10] of the carriers. Here, in this article, we have proposed a new three-level model in which a third upper level is introduced, with which the experimental results can be well simulated.

**RESULTS AND DISCUSSION**

The PL spectrum of a single QD excited by pulse laser with a repetition rate of $1 MHz$ is presented in Fig.1 (a), corresponding to the positively charged exciton ($X^+$, 908.15 nm), exciton (X, 911.25 nm) and biexciton emission (XX, 912.46 nm). These emission lines have been identified previously using the polarization-resolved PL spectra[11]. The emission intensities are greatly enhanced after the QD sample was transferred onto metal islands and the emission lines measured by corresponding PL and TRPL actually have a similar characteristics[6], thus we will focus on the emission dynamics of $X^+$ for analyzing QD emission with a long lifetime in detail. The typical TRPL spectrum is presented in Fig.1(b) at the condition of over-saturated excitation power $P = 2.5 \mu w$, showing a fast decay part with a lifetime of approximately 1 ns (see inset of Fig.1(b)) which can be fit by using a single exponential function and a very long lifetime decay of $100\ ns$ order of magnitude[6]. In addition, the corresponding $g^{(2)}(\tau)$ results at continuous wave laser excitation exhibits a bunching and antibunching characteristics, as shown in Fig.1(c). It is known that the antibunching property is related to a single QD emission, while the bunching property is generally related to the fact that there exists a third long lifetime shelving state below the excited state[12].

We note that the long lifetime decay curve cannot be fitted by a single exponential function. Actually, a similar non-single-exponential decay had been reported and interpreted by introducing a third long-lived level below the excited state of emission[7-10]. Here, in our case, a long lifetime decay curve is observed which is related to the fact that there exists a metastable state in the WL[6]. The energy level of WL is an upper level above the QD excited state. Thus, we propose a three-level model, as schematically shown in Fig.1(d), to deduce the exciton decay equations and simulate the experimental results. It is shown in Fig.1(d) that the QD emits from the excited state |2> to the ground state |1>, and the higher metastable state level |3> is the level of WL. This is a simplified model of the whole system. In fact, in the experiment, a 640 nm laser is used to excite the GaAs barrier at first and optically pumped electrons and holes will rapidly relax into the WL and QDs[5]. If there is no long lifetime metastable state, excitons in QDs and WL emit PL in a decay time of approximately 1ns. While if there exists a long lifetime metastable state in the WL, after initial exciton in QDs emits PL which corresponds to the fast decay process, more excitons in the WL will diffuse and be trapped by QDs again, which corresponds to the long lifetime decay process, as the measured TRPL curve representatively shown in Fig.1(b).

Based on the three-level model in Fig.1(d), we can establish a set of rate equation for describing the exciton emission from the QDs and WL after the optical excited electrons and holes relax to and populate the levels of |2> and |3>,

$$\frac{dn_2(t)}{dt} = n_3 r_{32} - n_2 r_{21}, \tag{1a}$$

$$\frac{dn_3(t)}{dt} = -n_3 r_{31} - n_3 r_{32} = -r' n_3, \tag{1b}$$

where $n_2$ and $n_3$ are the population rates of excitons in level |2> and |3>, $r_{12}$ and $r_{21}$ the excitation and radiation rates of level |2>, $r_{13}$ and $r_{31}$ the excitation and radiation rates of |3>, $r_{32}$ the trapped rate of exciton by QDs, and $r' \equiv r_{31} + r_{32}$. Next we will discuss the following two cases: ① level |3> is a normal state with a lifetime of approximately 1 ns and ② level |3> is a long lifetime metastable state.

For case ① $r_{32} \gg r_{21}, r_{31}$, Eq.(1) can be written as

$$n_3(t) = n_3(0)e^{-r't} \tag{2a}$$

$$\frac{dn_2(t)}{dt} = n_3(0)e^{-r't}r_{32} - n_2 r_{21} \tag{2b}$$

Setting $y \equiv n_2(t), a \equiv n_3(0)r_{32}, b \equiv r', c \equiv r_{21}$, Eq.(2b) can be rewritten as

$$\frac{dy}{dt} = ae^{-bt} - cy \tag{3}$$

The Laplace transformation is used to solve above equation, then we get the solution,

$$y(t) = y(0)e^{-ct} + \frac{a}{c-b}(e^{-bt} - e^{-ct}). \tag{4}$$

Thus, the population rate of energy level |2> is,

$$n_2(t) = n_2(0)e^{-r_{21}t} + \frac{n_3(0)r_{32}}{r_{21} - r'}\left(e^{-r't} - e^{-r_{21}t}\right) \tag{5}$$

Since $r_{32} \gg r_{21}, r_{31}$, QD emission intensity $I(t)$ can be expression as,

$$I(t) = r_{21}n_2(t) \propto (n_2(0)e^{-r_{21}t} + n_3(0)e^{-r_{21}t}) \propto e^{-r_{21}t}. \tag{6}$$

Thus Eq.(6) can be used to fit the fast decay part of TRPL curve, as shown in the inset of Fig.1b by red line. A lifetime of $\frac{1}{r_{21}} = 0.93 \pm 0.02\ ns$ is obtained. This result demonstrates that in this approximation the fluorescent decay lifetime of QDs is mainly determined by QD spontaneous radiation decay process.

For case ②, $r_{21} \gg r_{32}, r_{31}$, i.e., the energy level |3> is assumed to be a long lifetime

metastable state and QDs will emit luminescence after the excitons diffuse and are re-trapped by QDs. Hence, for the studied QD, the excitons nearby the QD will first relax into the QD and then emit the PL, leading to the formation of a depletion region near QDs. As time goes on, the depletion region expends, more and more excitons need to diffuse a larger distance and then be trapped by QD and emit PL. We can describe this process by dispersive diffusion with a characteristic power-law time dependence[13], i.e., the captured rate of QD is $r_{32} = r'_{32} t^{-\alpha}$, where $r'_{32}$ is a constant and the parameter $\alpha$ represents the dimension dependence[10]. In addition, it is assumed that $r_{31}$ and $r_{32}$ have the same order. It is approximated as $r_{31} + r_{32} \approx 2r_{32}$. Then the population rate of excitons in level |3> from the Eq.(1b) is,

$$n_3(t) = n_3(0) e^{-(rt)^\beta} \tag{7}$$

where $n_3(0)$ represents the population rate of excitons in the level |3> at $t = 0$, and $\beta = 1 - \alpha \in (0,1)$,[13] $r = (\frac{2r'_{32}}{1-\alpha})^{1/(1-\alpha)}$. Then based on Eq.(1a), the QD PL intensity of long lifetime decay part can be expressed as,

$$I(t) = r_{21} n_2(t) = n_3 r_{32} - \frac{dn_2(t)}{dt} \tag{8}$$

We consider the fact that the decay time of excitons in level |2> is approximately 1ns and it is assumed that $r_{21} \gg r_{32}$, then $\frac{dn_2(t)}{dt} \approx 0$ compared with the first term of $n_3 r_{32}$ in Eq.(8). Therefore, the PL intensity as a function of time is mainly determined by the trapping process of excitons, which can be written as,

$$I(t) \approx n_3 r_{32} = n_3(0) r'_{32} t^{\beta-1} e^{-(rt)^\beta} = A t^{\beta-1} e^{-(rt)^\beta} \tag{9}$$

where $A = n_3(0) r'_{32}$. This shows that the Eq.(9) has a stretched-like exponential decay characteristics and if $\beta = 1$ it becomes a single exponential decay function. From the expression of $I(t)$, we can derive the average decay time $<\tau>$ which can be written as,

$$<\tau> = \frac{\int_0^\infty t \cdot I(t) dt}{\int_0^\infty I(t) dt} = \frac{\int_0^\infty t^\beta e^{-(rt)^\beta} dt}{\int_0^\infty t^{\beta-1} e^{-(rt)^\beta} dt} = \frac{1}{r} \Gamma\left(\frac{1}{\beta} + 1\right) \tag{10}$$

where the $\Gamma$ is the mathematical Gamma function. Therefore, we can use the Eq.(9) to fit the TRPL curve, to get the fitting parameters of $r$ and $\beta$ and calculate the decay time $<\tau>$ from the Eq. (10). The fitting result using Eq.(9), as shown in Fig.(1b) by red line, gives $\frac{1}{r} = 194.4 \pm 0.6\ ns$ and $\beta = 0.876 \pm 0.002$. From the Eq.(10), the corresponding average decay time $<\tau> = 207.6 \pm 0.7\ ns$.

TRPL spectra as a function of excitation power are shown in Figs.2(a)-(c), where three typical TRPL results under excitation power of $P = 0.011, 0.132$ and $1.075\ \mu W$ are exhibited, corresponding to the long lifetime values of $\langle \tau \rangle = 590 \pm 20\ ns$, $303 \pm 2\ ns$ and $193.1 \pm 0.5\ ns$, respectively. These values are obtained by fitting the TRPL curves (red lines in Figs.2(a)-(c)) using Eq.(9) and calculating $\langle \tau \rangle$ using Eq.(10). The summarized plots of $<\tau>$ and the stretched factor $\beta$ as a function of excitation power are shown in Fig.2(d). It is shown that when excitation power increases the lifetimes $<\tau>$ decreases gradually from approximately 750 ns to 200 ns and $\beta$ changes from approximately 0.6 to 0.8. Since $\alpha = 1 - \beta$, α value will become smaller with increasing excitation power, implying that the trapping rate of excitons by QDs increases according to the expression of $r_{32} \sim t^{-\alpha}$ owing to the power-induced increase population rate of excitons in the WL.

Next we will analyze the measured $g^{(2)}(\tau)$ result shown in Fig.1(c), which exhibits a bunching and antibunching characteristics. Based on three-level model (see Fig.1(d)), and assuming that the population rate of excitons in level |1>, |2> and |3> is $p_1(t), p_2(t)$ and $p_3(t)$, respectively, the rate equation in matrix form can be written as,

$$\frac{d}{dt}\begin{pmatrix} p_1 \\ p_2 \\ p_3 \end{pmatrix} = -\begin{pmatrix} r_{12} + r_{13} & -r_{21} & -r_{31} \\ -r_{12} & r_{21} & -r_{32} \\ -r_{13} & 0 & r_{32} + r_{31} \end{pmatrix}\begin{pmatrix} p_1 \\ p_2 \\ p_3 \end{pmatrix} \qquad (11)$$

Its solution has the following form

$$\begin{pmatrix} p_1 \\ p_2 \\ p_3 \end{pmatrix} = \begin{pmatrix} a_{11} & a_{12} & a_{13} \\ a_{21} & a_{22} & a_{23} \\ a_{31} & a_{32} & a_{33} \end{pmatrix}\begin{pmatrix} \exp(-\lambda_1 t) \\ \exp(-\lambda_2 t) \\ \exp(-\lambda_3 t) \end{pmatrix}, \qquad (12)$$

where the $\lambda_i$ equals to the eigenvalues of the matrix of Eq.(12), i.e.,

$$\lambda_{1,2} = \frac{1}{2}\left[r_{12} + r_{13} + r_{21} + r_{32} + r_{31} \pm \sqrt{(r_{12} + r_{13} + r_{21} - r_{32} - r_{31})^2 - 4r_{13}r_{21} + 4r_{13}r_{31}}\right] (13a)$$

$$\lambda_3 = 0, \qquad (13b)$$

By considering the approximation of low transition rates to and from the metastable state |3>, $r_{13}$, $r_{31}$ and $r_{32}$ can be regarded as small quantities and $\lambda_1$ and $\lambda_2$ can be expressed as,

$$\lambda_1 \approx r_{12} + r_{21} \tag{14a}$$

$$\lambda_2 \approx r_{32} + r_{31} + \frac{r_{21}r_{13}}{r_{12} + r_{21}} \tag{14b}$$

$$\lambda_3 = 0. \tag{14c}$$

Considering the $t \to \infty$ limit, in which the system is in equilibrium, we can get

$$a_{23} = \frac{r_{12}r_{32} + r_{13}r_{32} + r_{12}r_{31}}{r_{13}r_{21} + r_{12}r_{31} + r_{12}r_{32} + r_{21}r_{31} + r_{21}r_{32} + r_{13}r_{32}} = p_2(\infty). \tag{15}$$

When $t = 0$, with the initial conditions $p_1(0) = 1, p_2(0) = p_3(0) = 0$ and the low transition rates approximation, we can get the expressions of the coefficients of $a_{21}$, $a_{22}$ and $a_{23}$ as follow,

$$a_{21} \approx \frac{-r_{12}}{r_{12} + r_{21}} \tag{16a}$$

$$a_{22} \approx \frac{r_{12}}{r_{12} + r_{21}} - \frac{r_{12}r_{32} + r_{13}r_{32} + r_{12}r_{31}}{r_{13}r_{21} + r_{12}r_{31} + r_{12}r_{32} + r_{21}r_{31} + r_{21}r_{32}} \tag{16b}$$

$$a_{23} \approx \frac{r_{12}r_{32} + r_{13}r_{32} + r_{12}r_{31}}{r_{13}r_{21} + r_{12}r_{31} + r_{12}r_{32} + r_{21}r_{31} + r_{21}r_{32}} \tag{16c}$$

Then the auto-correlation function $g^2(t)$ can be written as [14],

$$g^2(t) = \frac{p_2(t)}{p_2(\infty)} = \frac{a_{21}\exp(-\lambda_1 t) + a_{22}\exp(-\lambda_2 t) + a_{23}\exp(-\lambda_3 t)}{a_{23}} \tag{17}$$

By substituting the parameters in Eq.(17), $g^2(t)$ can be written as,

$$g^2(t) = 1 - (1 + a)\exp(-\lambda_1 t) + a\exp(-\lambda_2 t), \tag{18}$$

where $\lambda_1 = r_{12} + r_{21}$, $\lambda_2 = r_{31} + r_{32} + \frac{r_{21}r_{13}}{r_{12}+r_{21}}$ and $a = \frac{r_{21}r_{13}}{(r_{31}+r_{32})(r_{12}+r_{21})}$. It is worth noting that a similar expression of $g^2(t)$ function had been derived for different three-level configurations[12, 14-16] to fit the bunching and antibunching result, in which a third long-lived level located below the excited state was introduced. Here, in our case, the measured $g^2(t)$ result can be well fitted by using Eq.(18) if the t is taken as absolute value as shown in Fig.1(c) by red line, where the fitting parameters of $\frac{1}{\lambda_1} = 0.8 \pm 0.1\ ns$ and $\frac{1}{\lambda_2} = 172 \pm 9\ ns$. These short and long lifetime values have the same order as the decay lifetimes obtained by TRPL measurements. The fitting $g^2(0)$ value is

$0.23 \pm 0.09$.

**CONCLUSION**

In conclusion, we have proposed a three-level model to explain the observed long lifetime exciton emission in InAs/GaAs QD system. Assuming there exists a long lifetime metastable state in WL, the QD emission of long-lived part with non-single exponential decay in the TRPL spectra corresponds to a process of that the excitons in metastable state will diffuse and be re-trapped by QDs. From this model, we have derived a stretched-like exponential decay formula which can well describe the long lifetime decay curve. Furthermore, based on the proposed three-level model, we have derived an expression of $g^2(t)$ function which can be used to well fit the measured $g^2(t)$ result.

**METHODS**

The studied low density InAs/GaAs QD samples were grown by molecular beam epitaxy (MBE) on a (001) semi-insulating GaAs substrate. The corresponding discrete QD emission lines can be isolated by using confocal microscopy. Details of the sample growth procedure can be found elsewhere[17]. The epitaxial growth of InAs/GaAs QD samples consists of a 300 nm GaAs buffer layer, a 100 nm AlAs sacrificial layer, a 20 nm GaAs layer, an InAs WL of about one atomic layer thickness, an InAs QD layer, and a 100 nm GaAs cap layer. After etching away the AlAs sacrificed layer, the QD film of 120 nm was separated and transferred onto a polished metal sheet with random distribution of metal islands.

Photoluminescence (PL) and time-resolved photoluminescence (TRPL) spectra of QDs were performed at $7\ K$ using $640\ nm$ excitation from a pulsed semiconductor laser with a pulse length of $40\ ps$. For measuring the long-lifetime decay curves, the laser pulse repetition rate was set to $1 MHz$. A spot of about 2 $\mu m$ in diameter was focused on the sample using a confocal microscope objective lens (NA, 0.55). The PL spectral signal was extracted with the same microscope objective, dispersed by

500 $mm$ focal length monochromator and recorded by silicon charged coupled device (CCD). The TRPL spectra were measured by a time correlated single photon counting (TCSPC) device and the photon auto-correlation function $g^{(2)}(\tau)$ measurements were carried out using the Hanbury-Brown and Twiss (HBT) setup at 7 K using 640 nm excitation from a continuous wave laser.

## ASSOCIATED CONTENT

Corresponding Authors
*E-mail: xmdou04@semi.ac.cn
*E-mail: bqsun@semi.ac.cn

Notes
The authors declare no competing financial interest.

## ACKNOWLEDGMENTS

We acknowledge support from the National Key Research and Development Program of China (Grant No. 2016YFA0301202) and the National Natural Science Foundation of China (Grant Nos. 61827823 and 11974342).

## REFERENCES

(1) Purcell, E. M., SPONTANEOUS EMISSION PROBABILITIES AT RADIO FREQUENCIES. *Physical Review* **1946,** *69* (11-1), 681-681.

(2) Mie, G., Articles on the optical characteristics of turbid tubes, especially colloidal metal solutions. *Ann. Phys.-Berlin* **1908,** *25* (3), 377-445.

(3) García-Vidal, F. J.; Pendry, J. B., Collective Theory for Surface Enhanced Raman Scattering. *Phys Rev Lett* **1996,** *77* (6), 1163-1166.

(4) ElKabbash, M.; Miele, E.; Fumani, A. K.; Wolf, M. S.; Bozzola, A.; Haber, E.; Shahbazyan, T. V.; Berezovsky, J.; De Angelis, F.; Strangi, G., Cooperative Energy

Transfer Controls the Spontaneous Emission Rate Beyond Field Enhancement Limits. *Phys Rev Lett* **2019,** *122* (20), 203901.

(5) Drexhage, K. H., Storing light in the dark. *Prog. Optics* **1974,** *12*, 165.

(6) Chen, H.; Huang, J.; He, X.; Ding, K.; Ni, H.; Niu, Z.; Jiang, D.; Dou, X.; Sun, B., Plasmon-Field-Induced Metastable States in the Wetting Layer: Detected by the Fluorescence Decay Time of InAs/GaAs Single Quantum Dots. *ACS Photonics* **2020,** *7* (11), 3228-3235.

(7) Brosseau, C.-N.; Perrin, M.; Silva, C.; Leonelli, R., Carrier recombination dynamics inInxGa1−xN/GaNmultiple quantum wells. *Physical Review B* **2010,** *82* (8).

(8) Nikolaev, I. S.; Lodahl, P.; van Driel, A. F.; Koenderink, A. F.; Vos, W. L., Strongly nonexponential time-resolved fluorescence of quantum-dot ensembles in three-dimensional photonic crystals. *Physical Review B* **2007,** *75* (11).

(9) Sturman, B.; Podivilov, E.; Gorkunov, M., Origin of stretched exponential relaxation for hopping-transport models. *Phys Rev Lett* **2003,** *91* (17), 176602.

(10) Potuzak, M.; Welch, R. C.; Mauro, J. C., Topological origin of stretched exponential relaxation in glass. *J Chem Phys* **2011,** *135* (21), 214502.

(11) Dalgarno, P. A.; Smith, J. M.; McFarlane, J.; Gerardot, B. D.; Karrai, K.; Badolato, A.; Petroff, P. M.; Warburton, R. J., Coulomb interactions in single charged self-assembled quantum dots: Radiative lifetime and recombination energy. *Physical Review B* **2008,** *77* (24).

(12) Kurtsiefer, C.; Mayer, S.; Zarda, P.; Weinfurter, H., Stable solid-state source of single photons. *Phys. Rev. Lett.* **2000,** *85* (2), 290.

(13) Kakalios, J.; Street, R.; Jackson, n. W., Stretched-exponential relaxation arising from dispersive diffusion of hydrogen in amorphous silicon. *Phys. Rev. Lett.* **1987,** *59* (9), 1037.

(14) Kitson, S.; Jonsson, P.; Rarity, J.; Tapster, P., Intensity fluctuation spectroscopy of small numbers of dye molecules in a microcavity. *Physical Review A* **1998,** *58* (1), 620.

(15) Wu, E.; Jacques, V.; Zeng, H. P.; Grangier, P.; Treussart, F.; Roch, J. F., Narrow-band single-photon emission in the near infrared for quantum key distribution. *Optics Express* **2006,** *14* (3), 1296-1303.

(16) Reynaud, S., RESONANCE FLUORESCENCE - THE DRESSED ATOM APPROACH. *Annales De Physique* **1983,** *8* (4), 315-370.

(17) Yu, Y.; Shang, X.-J.; Li, M.-F.; Zha, G.-W.; Xu, J.-X.; Wang, L.-J.; Wang, G.-W.; Ni, H.-Q.; Dou, X.; Sun, B.; Niu, Z.-C., Single InAs quantum dot coupled to different "environments" in one wafer for quantum photonics. *Applied Physics Letters* **2013,** *102* (20).

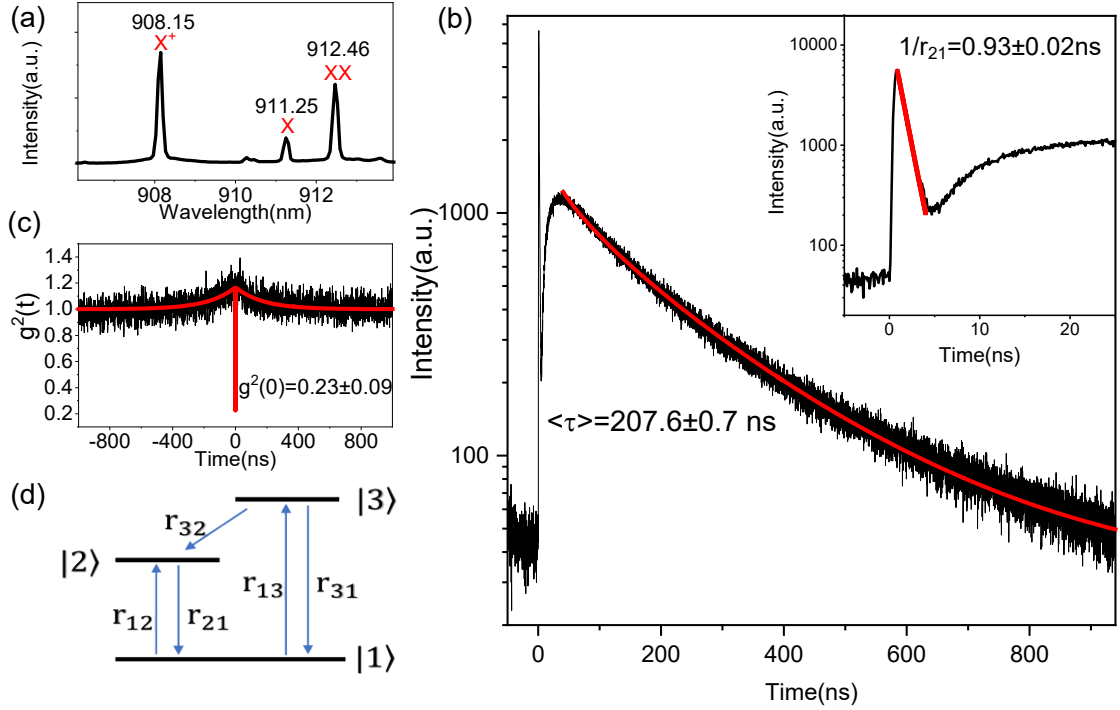

Fig.1 (a) PL spectrum of the positively charged exciton (X$^+$, 908.15 nm), exciton (X, 911.25 nm) and biexciton emission (XX, 912.46 nm). (b) TRPL of the X$^+$ emission at an excitation of power of 2.5 $\mu$W. The red line is a fitting result by using Eq.(9) and obtained average lifetime of $<\tau>= 207.6 \pm 0.7\ ns$. Inset: Zoomed in at 0-20 ns of TRPL and linear fitting to the data (red line) with a decay time of $0.93 \pm 0.02\ ns$. (c) $g^2(t)$ measurement of the QD emission with continuous wave excitation at an excitation power of 0.043 $\mu$W. The red line is a fitting result by using Eq.(18). The $g^2(0)$ value is obtained as $0.23 \pm 0.09$. (d) Schematic diagram of three-level model with a long-lived metastable level |3> and an excited state of QD |2> as well as the ground state |1>.

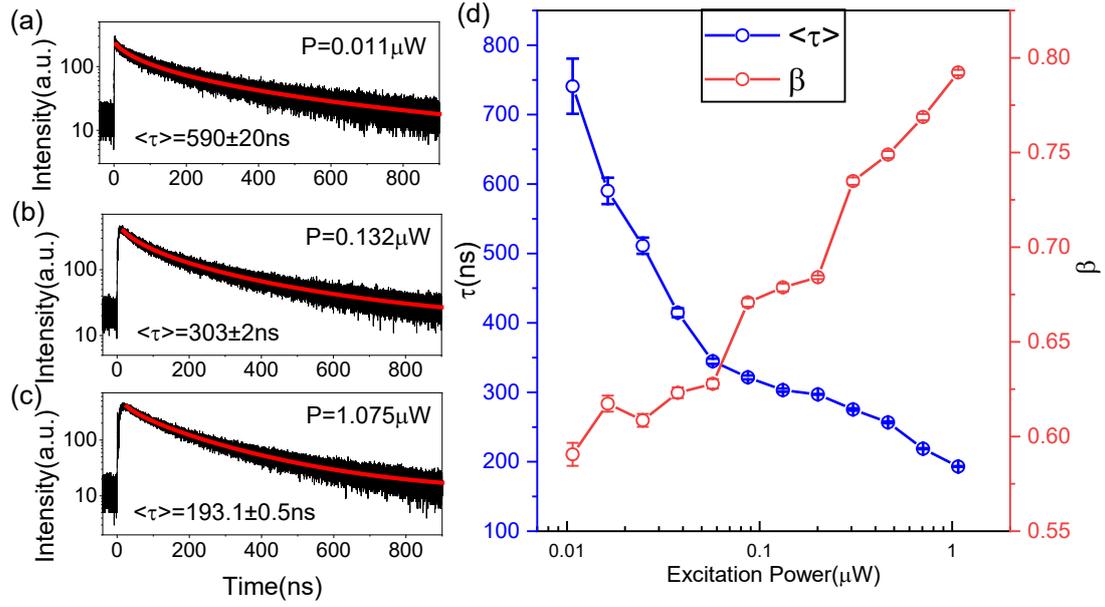

Fig.2 (a)-(c) TRPL spectra at the excitation powers of 0.011, 0.132 and 1.075 $\mu$W, the red lines are the fitting results by using Eq.(9) and the obtained average lifetimes are $590 \pm 20$, $303 \pm 2$ and $193.1 \pm 0.5\ ns$, respectively. (d) Excitation power dependence of the lifetime $\tau$ and fitting parameter $\beta$ obtained by using the Eqs.(9) and (10).